\documentclass{revtex4}
\usepackage{epsfig}
\usepackage{amssymb}
\usepackage{amsmath}
\usepackage{amsfonts}
\usepackage{graphicx}
\usepackage{mathrsfs}
\usepackage[dvips]{color}
\usepackage{multirow}

\newcommand{\R}{\mathbb{R}}
\newcommand{\C}{\mathbb{C}}

\newcommand{\fa}{\mathfrak{a}}
\newcommand{\fb}{\mathfrak{b}}

\newcommand{\fn}{{\,\mathfrak{n}\,}}

\newcommand{\fz}{\mathfrak{z}}

\newcommand{\bH}{\mathbf{H}}

\newcommand{\bM}{\mathbf{M}}
\newcommand{\bN}{\mathbf{N}}

\newcommand{\bU}{\mathbf{U}}

\newcommand{\bsigma}{\mathbf{\sigma}}

\newcommand{\cO}{\mathcal{O}}
\newcommand{\cP}{\mathcal{P}}

\newcommand{\cT}{\mathcal{T}}

\newcommand{\be}{\begin{equation}}
\newcommand{\ee}{\end{equation}}
\newcommand{\bea}{\begin{eqnarray}}
\newcommand{\eea}{\end{eqnarray}}
\newcommand{\nn}{\nonumber}

\newcommand{\ed}{\end{document}}

\newcommand{\bi}{\begin{itemize}}
\newcommand{\ei}{\end{itemize}}

\newcommand{\bce}{\begin{center}}
\newcommand{\ece}{\end{center}}

\newcommand{\sH}{\mathscr{H}}

\newcommand{\sT}{\mathscr{T}}
\newcommand{\sU}{\mathscr{U}}

\newcommand{\nuj}{j}

\begin{document}

\title{Transfer Matrices as Non-Unitary $S$-Matrices, Multimode  Unidirectional\\ Invisibility, and Perturbative Inverse Scattering}

\author{Ali~Mostafazadeh}
\address{Departments of Physics and Mathematics, Ko\c{c}
University,\\ Sar{\i}yer 34450, Istanbul, Turkey\\
amostafazadeh@ku.edu.tr}

\begin{abstract}
We show that in one dimension the transfer matrix $\bM$ of any scattering potential $v$ coincides with the $S$-matrix of an associated time-dependent non-Hermitian $2\times 2$ matrix Hamiltonian $\bH(\tau)$. If $v$ is real-valued, $\bH(\tau)$ is pseudo-Hermitian and its exceptional points correspond to the classical turning points of $v$. Applying time-dependent perturbation theory to $\bH(\tau)$ we obtain a perturbative series expansion for $\bM$ and use it to study the phenomenon of unidirectional invisibility. In particular, we establish the possibility of having multimode unidirectional invisibility with wavelength-dependent direction of invisibility and construct various physically realizable optical potentials possessing this property. We also offer a simple demonstration of the fact that the off-diagonal entries of the first Born approximation for $\bM$ determine the form of the potential. This gives rise to a perturbative inverse scattering scheme that is particularly suitable for optical design. As a simple application of this scheme, we construct an infinite-range unidirectionally invisible potential.\\

\noindent {Pacs numbers: 03.65.Nk, 42.25.Bs, 02.30.Zz}\\


\end{abstract}

\maketitle

\section{Introduction}

Recently there has been a growing research activity \cite{prl-2009} -- \cite{pra-2013a} in the study of complex scattering potentials in one dimension \cite{muga}. Unlike the real scattering potentials, these can support spectral singularities \cite{prl-2009} -- \cite{CG} which provide the mathematical basis for lasing \cite{pra-2011a}, the time-reversal spectral singularities \cite{longhi} which are responsible for antilasing \cite{anti-laser}, and unidirectional reflectionlessness and invisibility \cite{invisible1} -- \cite{pra-2013a,invisible-old} which are of great interest for applications in optical circuitry \cite{invisible3,invisible4} and cold atom physics \cite{rmr}. A particularly useful tool in the study of these phenomena is the transfer matrix \cite{sanchez}. In the present article we use a two-component formulation of the scattering problem to show that the transfer matrix of a general, possibly complex and energy-dependent, scattering potential coincides with the $S$-matrix of a particular non-Hermitian two-level Hamiltonian. This observation allows for a thorough perturbative analysis of the transfer matrix, elucidates some of the interesting aspects of the inverse scattering theory, and paves the way for the construction of a remarkable class of physically realizable optical potentials that display multimode unidirectional invisibility with wavelength-dependent direction of invisibility.

The use of two-component formulations of field equations to study Klein-Gordon and similar fields has a long history \cite{kg-old} -- \cite{cqg-2003-ap-2004}. In Ref.~\cite{jmp-1998} we use a particular type of two-component state vectors to study the solutions of the Wheeler-DeWitt equation for a certain quantum cosmological model. These play a central role in the study of relativistic and cosmological geometric phases \cite{jpa-1998b-tjp-1999} and the development of a consistent quantum mechanics of Klein-Gordon-type fields \cite{cqg-2003-ap-2004}. The latter admits a generalization for treating first quantized massive spin 1 (Proca) fields \cite{jmp-2009}. A mathematically less complicated but physically more relevant task is to devise a two-component formulation of the time-independent Schr\"odinger equation. The authors of Ref.~\cite{U-M} undertake this task to study a class of complex scattering potentials that support unidirectional invisibility. In Ref.~\cite{p114} we use a different type of two-component state vectors to develop a dynamical formulation of one-dimensional scattering theory where the reflection and transmission amplitudes are given as solutions of a set of dynamical equations.  In the present article, we employ the two-component state vectors employed in Refs.~\cite{jmp-1998,jpa-1998b-tjp-1999,cqg-2003-ap-2004} to reveal some conceptually and practically useful aspects of the formulation of scattering theory in terms of the transfer matrix. These have direct applications in designing optical potentials with appealing properties and should be useful in the study of a variety of condensed matter and atomic physics systems that admit a description in terms of the transfer-matrix approach to scattering theory \cite{muga,sanchez}.

In the reminder of this section, we give the definition of the transfer matrix and summarize its basic properties.

Consider the time-independent Schr\"odinger equation,
	\be
	-\psi''(x)+v(x)\psi=k^2\psi(x),
	\label{sch}
	\ee
where $x\in\R$ and $v$ is a possibly complex and energy-dependent scattering potential; so that $v(x)\to 0$ as $x\to\pm\infty$. The scattering solutions of (\ref{sch}) correspond to real and positive values of the wavenumber $k$ that satisfy
	\begin{align}
	&\psi_k^l(x)=\left\{
	\begin{array}{ccc}
	e^{ikx}+R^l e^{-ikx} & {\rm for} & x\to-\infty,\\
	T e^{ikx}& {\rm for} & x\to\infty,
	\end{array}\right.&&
	\psi_k^r(x)=\left\{ \begin{array}{ccc}
	T e^{-ikx} & {\rm for} & x\to-\infty,\\
	e^{-ikx}+R^r e^{ikx}& {\rm for} & x\to\infty.
	\end{array}\right.
	\label{scatter}
	\end{align}
Here $R^l$, $R^r$, and $T$ are respectively the left-reflection, right-reflection, and transmission amplitudes.

The general solution of (\ref{sch}) has the asymptotic form:
	\be
	\psi(x)=A_\pm e^{ikx}+B_\pm e^{-ikx}~~~{\rm for}~~~x\to\pm\infty,
	\label{asym}
	\ee
where $A_\pm$ and $B_\pm$ are complex coefficients possibly depending on $k$. The transfer matrix associated with $v$ is the unique $2\times 2$ matrix $\bM$ fulfilling
	\be
	\left[\begin{array}{c}
	A_+\\ B_+\end{array}\right]=\bM\left[\begin{array}{c}
	A_-\\ B_-\end{array}\right].
	\label{M}
	\ee
It has the following  useful properties \cite{muga,jpa-2009,prl-2009,sanchez}.
	\begin{enumerate}
	\item $\bM$ has a unit determinant.

	\item $\bM$ encodes all the information about the scattering properties of $v$. In particular, its entries $M_{ij}$ are related to the reflection and transmission amplitudes according to
		\begin{align}
		&M_{11}=T-\frac{R^l R^r}{T^2}, &&
		M_{12}=\frac{R^r}{T}, && M_{21}=-\frac{R^l}{T}, && M_{22}=\frac{1}{T}.
		\label{M-RT}
		\end{align}
	\item The bound states and resonances of $v$ correspond to complex values of $k$ for which 	
	$M_{22}$ vanishes.
	\item $\bM$ has a simple composition property in the following sense. Suppose that $v_1$ and
	$v_2$ are potentials vanishing outside a pair of adjacent intervals $[a_1,a_2]$ and
    $[a_2,a_3]$ in $\R$, and $v=v_1+v_2$. If $\bM_1$, $\bM_2$, and $\bM$ are respectively the transfer matrices for $v_1$, $v_2$, and $v$, we have $\bM=\bM_2\bM_1$.
 	\end{enumerate}

\section{A Two-Component Formulation of Scattering}
\label{sec2}

Let $\tau:=kx$ and $\phi(\tau):=\psi(\tau/k)$. Then we can express the time-independent Schr\"odinger equation~(\ref{sch}) in the form of the time-dependent Schr\"odinger equation,
	\be
	i\dot\Psi(\tau)=\bH(\tau)\Psi(\tau),
	\label{sch-eq}
	\ee
where an over-dot represents a derivative with respect to $\tau$,
	\begin{align}
	\Psi(\tau)&:=\frac{1}{2}\left[\begin{array}{c}
	\phi(\tau)-i\dot\phi(\tau)\\
	\phi(\tau)+i\dot\phi(\tau)\end{array}\right],
	\label{def1}\\
	\bH(\tau)&:=\left[\begin{array}{cc}
	w(\tau)-1 & w(\tau)\\
	-w(\tau) & -w(\tau)+1\end{array}\right]
	=-\bsigma_3+w(\tau)\bN,
	\label{def2}\\
	w(\tau)&:=\frac{v(\tau/k)}{2k^2},~~~~~~~~~
    \bN:=i\bsigma_2+\bsigma_3=
    \left[\begin{array}{cc} 1 & 1 \\-1 & -1\end{array}\right],
	\end{align}
and $\bsigma_i$, with $i=1,2,3$, are Pauli matrices. Alternatively, for all $\tau,\tau_0\in\R$,
	\be
	\Psi(\tau)=\bU(\tau,\tau_0)\Psi(\tau_0),
	\label{U=}
	\ee
where $\bU(\tau,\tau_0):=\sT e^{-i\int_{\tau_0}^\tau\bH(\tau')d\tau'}$ is the time-evolution operator associated with $\bH(\tau)$, and $\sT$ is the time-ordering operator.

Notice that the two-component state vector $\Psi$ and the two-level Hamiltonian $\bH(\tau)$ are essentially identical with those used in Refs.~\cite{jmp-1998,jpa-1998b-tjp-1999,cqg-2003-ap-2004}. In particular, whenever $v$ and subsequently $w$ are real-valued, we have $\bH(\tau)^\dagger=\bsigma_3^{-1}\bH(\tau)\bsigma_3$. This shows that $\bH(\tau)$ is a $\bsigma_3$-pseudo-Hermitian Hamiltonian operator with real or complex-conjugate pair of eigenvalues \cite{p123}. Indeed the eigenvalues of $\bH(\tau)$ are give by $E_\pm(\tau):=\pm\fn(\tau)$, where
    \be
    \fn(\tau):=\sqrt{1- 2 w(\tau)}=\sqrt{1-\frac{v(\tau/k)}{k^2}}.
    \ee
Therefore, as we expect, they are either real or form a complex-conjugate pair for real-valued potentials $v$. These respectively happen whenever $v(x)\leq k^2$ and
$v(x)>k^2$. If $x$ is a classical turning point, i.e., $v(x)=k^2$, $E_\pm(\tau)$ coincide and vanish identically, and $\bH(\tau)$ becomes non-diagonalizable. This shows that the classical turning points correspond to the exceptional points of $\bH(\tau)$, \cite{eps,jmp-2008}. In particular, tunnelling through a real barrier potential corresponds to time-evolutions in which the parameters of $\bH(\tau)$ pass through the exceptional points. Outside the barrier, $\bH(\tau)$ is a diagonalizable matrix with real eigenvalues, and as a result it is quasi-Hermitian \cite{SGH,review}. Inside the barrier, it is pseudo-Hermitian but not quasi-Hermitian \endnote{Note also that $\bH(\tau)$ is a real matrix, i.e., it commutes with the time-reversal operator $\cT$ defined by $\cT\Phi:=\Phi^*$, where $\Phi$ is an arbitrary two-component state vector.}.

Next, let $\tau_\pm\in[-\infty,\infty]$ be such that $(\tau_-,\tau_+)$ is the largest open interval outside which $w$ vanishes identically, and
	   \be
	   \bU_0(\tau):=e^{i\bsigma_3 \tau}=\left[\begin{array}{cc}
	   e^{i\tau}& 0 \\
	   0 & e^{-i\tau}\end{array}\right],~~~~\tau\in\R.
	   \label{q2}
	   \ee
Then, for all $\tau\leq \tau_-$, $\psi(x)=A_-e^{ikx}+B_-e^{-ikx}$ for some $A_-,B_-\in\C$, and in light of (\ref{def1}) we have
        	\be
        	\Psi(\tau)=\left[\begin{array}{c} A_-e^{i\tau}\\B_-e^{-i\tau} \end{array}\right]=
        	\bU_0(\tau)\left[\begin{array}{c} A_-\\B_- \end{array}\right],~~~~\tau\leq \tau_-.
        	\label{q3}
        	\ee
Similarly, for all $\tau\geq \tau_+$, $\psi(x)=A_+e^{ikx}+B_+e^{-ikx}$ for some $A_+,B_+\in\C$, and
         	\be
        	\Psi(\tau)=\left[\begin{array}{c} A_+e^{i\tau}\\B_+e^{-i\tau} \end{array}\right]=
        	\bU_0(\tau)\left[\begin{array}{c} A_+\\ B_+ \end{array}\right],~~~~\tau\geq \tau_+.
        	\label{q4}
        	\ee
According to (\ref{U=}), we also have
	\be
	\Psi(\tau_+)=\bU(\tau_+,\tau_-)\Psi(\tau_-).
	\label{q5}
	\ee
Inserting $t=\tau_-$ and $t=\tau_+$ respectively in (\ref{q3}) and (\ref{q4}), and using the resulting expressions and (\ref{q5}), we find
	\be
	\bU_0(\tau_+)\left[\begin{array}{c} A_+\\ B_+ \end{array}\right]=\bU(\tau_+,\tau_-)
	\bU_0(\tau_-)\left[\begin{array}{c} A_-\\ B_- \end{array}\right].
	\ee
Applying $\bU_0(\tau_+)^{-1}$ to both sides of this equation and comparing the result with (\ref{M}), we obtain
	\be
	\bM=\bU_0(\tau_+)^{-1}\bU(\tau_+,\tau_-)\bU_0(\tau_-).
	\label{M-2}
	\ee
For a finite-range potential, $\tau_\pm$ are real numbers and we can use this equation directly. For an infinite-range potential, Eq.~(\ref{M-2}) is meaningful asymptotically and we interpret it as
	\be
    	\bM=\lim_{\tau_\pm\to\pm\infty}\bU_0(\tau_+)^{-1}\bU(\tau_+,\tau_-)\bU_0(\tau_-)=:
	\bU_0(\infty)^{-1}\bU(\infty,-\infty)\bU_0(-\infty).
	\label{M-3}
	\ee
This relation holds also for finite-range potentials, because they vanish outside $[\tau_-,\tau_+]$.
	
In view of (\ref{q2}), we can identify $\bU_0(\tau)$ with $\bU_0(\tau,0)$, where $\bU_0(\tau,\tau_0):=e^{-i(\tau-\tau_0)\bH_0}=e^{i(\tau-\tau_0)\sigma_3}$ is the time-evolution operator for the Hamiltonian $\bH_0:=-\bsigma_3$, which describes a free particle. In view of this observation and the definition of the $S$-matrix \cite[\S 3.2]{weinberg}, Eq.~(\ref{M-3})  implies the following remarkable result.
    \begin{itemize}
    \item[]\textbf{Theorem~1:} Let $v$ be a possibly complex and energy-dependent scattering potential defined on the real line and $\bH(\tau)$ be the two-level Hamiltonian given by (\ref{def2}). Then the transfer matrix of $v$ coincides with the $S$-matrix of $\bH(\tau)$.
	\end{itemize}

\section{Perturbative Expansion of the Transfer Matrix}
	
The particular form of Eq.~(\ref{M-2}) suggests that we can identify $\bM$ with the asymptotic value of the evolution operator for the two-level system obtained by performing the time-dependent unitary transformation: $\Psi(\tau)\to \bU_0(\tau)^{-1}\Phi(\tau)$. This is precisely the transformation to the interaction picture \cite{sakurai} where the Hamiltonian and the time-evolution operator take the following form, respectively.
	\begin{align}
	\sH(\tau)&:=\bU_0(\tau)^{-1}\bH(\tau) \bU_0(\tau)-i\bU_0(\tau)^{-1}\dot \bU_0(\tau)
	\nn\\
	&=w(\tau)\,
	e^{-i\bsigma_3 \tau} \bN e^{i\bsigma_3 \tau}=
	w(\tau)\,\left[\begin{array}{cc}
	1 & e^{-2i\tau}\\
	-e^{2i \tau} & -1\end{array}\right],
	\label{int-H}\\
	\sU(\tau,\tau_0)&:=\bU_0(\tau)^{-1}\bU(\tau,\tau_0)\bU_0(\tau_0).
	\label{int-U}
	\end{align}
According to Eqs.~(\ref{M-2}) -- (\ref{int-U}), $\sH(\tau)$ is a non-diagonalizable time-dependent Hamiltonian whose evolution operator $\sU(\tau,\tau_0)$ satisfies $\sU(\tau_+,\tau_-)=\bM$. In particular, we have \cite{p114}
    \be
	\bM=\sU(\infty,-\infty).
	\label{M-5}
	\ee

Because $\sH(\tau)$ is proportional to $w(\tau)$, we can use (\ref{M-5}) to obtain the following perturbative expansion for $\bM$, \cite{sakurai}.
    \be
    \bM=\sT e^{-i\int_{-\infty}^\infty d\tau \sH(\tau)}=1+\sum_{\ell=1}^\infty \bM^{(\ell)},
    \label{perturb}
    \ee
where
    {\bea
    \bM^{(\ell)}&:=&(-i)^\ell\int_{-\infty}^\infty d\tau_\ell \int_{-\infty}^{\tau_\ell} d\tau_{\ell-1}\cdots
    \int_{-\infty}^{\tau_2} d\tau_1 \sH(\tau_\ell)\sH(\tau_{\ell-1})\cdots\sH(\tau_1)\nn\\
    &=& (-i)^\ell \int_{-\infty}^\infty d\tau_\ell \int_{-\infty}^{\tau_\ell} d\tau_{\ell-1}\cdots\int_{-\infty}^{\tau_2} d\tau_1
    e^{-i\tau_\ell\bsigma_3}\bN e^{i(\tau_\ell-\tau_{\ell-1})\bsigma_3}\bN\cdots
    e^{i(\tau_2-\tau_1)\bsigma_3}\bN e^{i\tau_1\bsigma_3} \prod_{p=1}^\ell w(\tau_p) \nn\\
    &=&\frac{(-i)^\ell}{2^\ell k^{\ell}}\int_{-\infty}^\infty dx_\ell \int_{-\infty}^{x_\ell} dx_{\ell-1}\cdots\int_{-\infty}^{x_2} dx_1 \prod_{p=1}^\ell v(x_p) \Big\{
    e^{-ikx_\ell\bsigma_3}\bN \times \nn\\
    && \hspace{5cm}
     e^{ik(x_\ell-x_\ell)\bsigma_3}\bN\cdots
    e^{ik(x_2-x_1)\bsigma_3}\bN e^{ik x_1\bsigma_3}\Big\}.
    \label{series}
    \eea}
In particular, we have
    \bea
    \bM^{(1)}&=&\frac{-i}{2k}\left[\begin{array}{cc}
    \tilde v(0) & \tilde v(2k)\\[3pt]
    -\tilde v(-2k) & -\tilde v(0)\end{array}\right],
    \label{M1=}\\
    \bM^{(2)}&=&\frac{-1}{4k^2}\left[\begin{array}{cc}
    \tilde v(0,0)-\tilde v(-2k,2k) & \tilde v(2k,0)-\tilde v(0,2k)\\[3pt]
    \tilde v(-2k,0)-\tilde v(0,-2k) & \tilde v(0,0)-\tilde v(2k,-2k)\end{array}\right],
    \label{M2=}
    \eea
where for every function $f$ of $x_1,x_2,\cdots,x_\ell\in\R$, $\tilde f$ is the Fourier transform of $f$ that is given by
    \be
    \tilde f(k_1,k_2,\cdots,k_\ell):=\int_{-\infty}^\infty dx_1\int_{-\infty}^\infty dx_2
    \cdots \int_{-\infty}^\infty dx_\ell\:
    e^{-i(k_1x_1+\cdots+k_\ell x_\ell)} f(x_1,x_2,\cdots,x_\ell),
    \label{Fourier}
    \ee
$v(x_1,x_2):=v(x_2)\theta(x_2-x_1)v(x_1)$, and $\theta(x)$ stands for the step function defined by
    \be
    \theta(x):=\left\{\begin{array}{ccc}
    0&{\rm for}& x\leq 0,\\
    1&{\rm for}& x>0.\end{array}\right.
    \ee

According to (\ref{perturb}) -- (\ref{M1=}), the (first) Born approximation yields
    \be
    \bM\approx 1+\bM^{(1)}=\left[\begin{array}{cc}
    1-i\tilde v(0)/2k & -i\tilde v(2k)/2k\\[3pt]
    i\tilde v(-2k)/2k & 1+i\tilde v(0)/2k\end{array}\right].
    \label{M-pert-1}
    \ee
In view of (\ref{M-RT}), this implies
    \begin{align}
    &R^l \approx R^{l}_1(k):=\frac{\tilde v(-2k)}{2i k-\tilde v(0)}, &&
    R^r \approx R^{r}_1(k):= \frac{\tilde v(2k)}{2i k-\tilde v(0)}, &&
    T \approx  T_1(k):=\frac{2ik}{2ik-\tilde v(0)}.
    \label{RRT-pert-1}
    \end{align}
Similarly, second-order perturbation theory, i.e., $\bM\approx 1+\bM^{(1)}+\bM^{(2)}$, gives
    \begin{align}
    & R^l \approx R^l_2(k):=\frac{-2ik\,\tilde v(-2k)+\tilde v(-2k,0)-\tilde v(0,-2k)}{4k^2+2i\tilde v(0)\,k+\tilde v(2k,-2k)-\tilde v(0,0)},
    \label{RL-2nd}\\[6pt]
    & R^r \approx R^r_2(k):=\frac{-2ik\,\tilde v(2k)-\tilde v(2k,0)+\tilde v(0,2k)}{4k^2+2i\tilde v(0)\,k+\tilde v(2k,-2k)-\tilde v(0,0)},
    \label{RR-2nd}\\[3pt]
    &T \approx T_2(k):=\frac{4k^2}{4k^2+2i\tilde v(0)\,k+\tilde v(2k,-2k)-\tilde v(0,0)}.
    \label{T-2nd}
    \end{align}

As a simple example consider the double-delta-function potential \cite{jpa-2009},
    \be
    v(x)=\fz_1\delta(x-a_1)+\fz_2\delta(x-a_2),
    \label{double-d}
    \ee
where $\fz_1$ and $\fz_2$ are possibly complex coupling constants, and $a_1$ and $a_2$ are real parameters. As shown in Ref.~\cite{pra-2012}, the second order perturbation theory gives the exact value of the transfer matrix for this potential. The calculation of $\tilde v(k)$ and $\tilde v(k_1,k_2)$ for (\ref{double-d}) is straightforward. The result is
	\begin{align}
	&\tilde v(k)=\fz_1 e^{-ika_1}+\fz_2 e^{-ika_2}, &&
	\tilde v(k_1,k_2)=\fz_1\fz_2\left[e^{-i(k_1a_2+k_2a_1)}\theta(a_1-a_2)+
	e^{-i(k_1a_1+k_2a_2)}\theta(a_2-a_1)\right].\nn
	\end{align}
Substituting these relations in (\ref{M1=}) and (\ref{M2=}) and using (\ref{perturb}), we find
	\bea
	M_{11}(k)=M_{22}(-k)&=&1-\frac{i(\fz_1+\fz_2)}{2k}-\frac{\fz_1\fz_2}{4k^4} \left[1-e^{2ik(a_2-a_1)}\theta(a_1-a_2)-e^{2ik(a_1-a_2)}\theta(a_2-a_1)\right],\nn\\
	M_{12}(k)=M_{21}(-k)&=& -\frac{i}{2k}\left(\fz_1e^{-2ika_1}+\fz_2e^{-2ika_2}\right)\nn\\
	&&-\frac{\fz_1\fz_2}{4k^4}
	\left[\left(e^{-2ika_2}-e^{-2ika_1}\right)\theta(a_1-a_2)+
	\left(e^{-2ika_1}-e^{-2ika_2}\right)\theta(a_2-a_1)\right],\nn
	\eea
which agree with the results reported in Refs.~\cite{jpa-2009,pra-2012}.

\section{Perturbative and Multimode Unidirectional Invisibility}

Consider the complex potential
    \be
    v(x)=\left\{\begin{array}{cc}
    \fz\,e^{iK x}&{\rm for}~x\in[0,L],\\
    0 &{\rm otherwise},\end{array}\right.
    \label{exp-v}
    \ee
where $\fz$ is a possibly $k$-dependent coupling constant and $K$ is a characteristic wavenumber \cite{invisible1,invisible,U-M,invisible-old}. This is a finite-range locally periodic potential \cite{griffiths} whose scattering properties can be easily explored using our perturbative formulas for the reflection and transmission amplitudes. This potential is of particular interest, because it provides a physically realizable toy model \cite{invisible2} displaying unidirectional invisibility \cite{invisible1,pra-2013a,invisible-old}. Its infinite-range analog, namely $v(x)=\fz\,e^{iK x}$ for all $x\in\R$, is studied in Refs.~\cite{gasymov,exp-pot}.

Substituting (\ref{exp-v}) in (\ref{RRT-pert-1}) and carrying out the necessary calculations, we find
    \bea
    R^l&=&-\frac{\left[e^{i(2k+K)L}-1\right]\fz}{2k(2k+K)}+\cO(\fz^2),
    \label{exp-RL-1}\\
    R^r&=&\frac{\left[e^{-i(2k-K)L}-1\right]\fz}{2k(2k-K)}+\cO(\fz^2),
    \label{exp-RR-1}\\
    T&=&1+\frac{i\left(e^{iK L}-1\right)\fz}{2K k}+\cO(\fz^2),
    \label{exp-T-1}
    \eea
where $\cO(\fz^d)$ stands for the terms of order $d$ and higher in powers of $\fz$.

According to (\ref{exp-RL-1}), for small values of the coupling constant where we can ignore the quadratic and higher order terms in $\fz$, the potential (\ref{exp-v}) is reflectionless from the left provided that $(2k+K )L$ is an integer multiple of $2\pi$. Furthermore, according to (\ref{exp-T-1}), it is invisible from the left in the sense that $R^l=\cO(\fz^2)=T-1$, if $K L$ is also an integer multiple of $2\pi$, i.e.,
    \be
    (2k+K )L=2\pi n,~~~~~~~~K L=2\pi m,~~~~~~~~~m,n=1,2,3,\cdots.
    \label{condi-1}
    \ee
However, it is easy to check that this invisibility is not unidirectional unless $n=2m$. In view of (\ref{condi-1}), this corresponds to
    \be
    k=\frac{K }{2}=\frac{2\pi m}{L}.
    \label{condi-2}
    \ee
Under this condition we have unidirectional invisibility from the left, if we can ignore the quadratic and higher order terms in $\fz$. We refer to this property as ``perturbative unidirectional invisibility.'' Similarly we can speak of ``perturbative unidirectional reflectionlessness.''

We can use (\ref{RL-2nd}) -- (\ref{T-2nd}) to compute the contribution of the quadratic terms to the reflection and transmission amplitudes. Enforcing (\ref{condi-2}), we then find
    \begin{align}
    &R^l=\cO(\fz^3), && R^r=\frac{-iL^2\fz}{4\pi m}+\frac{iL^4\fz^2}{32\pi^3 m^3}
    +\cO(\fz^3), &&T=1+\frac{iL^4\fz^2}{128\pi^3 m^3}+\cO(\fz^3).
    \label{uni-dir-2}
    \end{align}
Therefore, under the condition (\ref{condi-2}) and up to and including the quadratic terms in $\fz$, the potential (\ref{exp-v}) is reflectionless but not invisible from the left. This is in complete agreement with the results of Refs.~\cite{invisible}.

Next, consider a general finite-range locally periodic potential of the form
    \be
    v(x)=\left\{\begin{array}{cc}
    \fz\,f(x)&{\rm for}~x\in[0,L],\\
    0 &{\rm otherwise},\end{array}\right.
    \label{LP-v}
    \ee
where $f$ is a piecewise continuous periodic function with period $\ell$, i.e., $f(x+\ell)=f(x)$ for all $x\in\R$. Expanding $f(x)$ in its complex Fourier series in $[0,\ell]$, i.e., setting
$f(x)=\sum_{\nuj=-\infty}^\infty c_\nuj e^{i \nuj K  x}$ with
$c_\nuj:=\ell^{-1}\int_0^\ell e^{-i\nuj K x}v(x)dx$ and $K:= 2\pi/\ell$, we can express (\ref{LP-v}) as
    \be
    v(x)=\sum_{\nuj=-\infty}^\infty v_\nuj(x),
    \label{LP-v2}
    \ee
where $v_\nuj(x)$ is a potential of the form (\ref{exp-v}) with $\fz\to \fz\,c_\nuj$ and
$K\to K\nuj$.

Equations (\ref{RRT-pert-1}) and (\ref{LP-v2}) suggest that we can read out the result of the first-order perturbative calculation of the reflection and transmission amplitudes for the potential (\ref{LP-v}) from that of (\ref{exp-v}). In this way we obtain, using (\ref{exp-RL-1}) -- (\ref{exp-T-1}),
    \bea
    R^l&=&-\left[\frac{c_0(e^{2ikL}-1)}{2k}+\sum_{\nuj=1}^\infty\left\{
    \frac{c_{-\nuj}\left[e^{i(2k-\nuj K)L}-1\right]}{2k-\nuj K}+
    \frac{c_\nuj\left[e^{i(2k+\nuj K)L}-1\right]}{2k+\nuj K}\right\}\right]\frac{\fz}{2k}+\cO(\fz^2),\nn\\
    R^r&=&\left[\frac{c_0(e^{-2ikL}-1)}{2k}+\sum_{\nuj=1}^\infty\left\{
    \frac{c_{-\nuj}\left[e^{-i(2k+\nuj K)L}-1\right]}{2k+\nuj K}+
    \frac{c_\nuj\left[e^{-i(2k-\nuj K)L}-1\right]}{2k-\nuj K}\right\}\right]\frac{\fz}{2k}+\cO(\fz^2),\nn\\
    T&=&1-\left[c_0 L+\frac{i}{K}\sum_{\nuj=1}^\infty\frac{1}{\nuj} \Big\{
    c_{-\nuj}\left[e^{-i\nuj KL}-1\right] - c_\nuj\left[e^{i\nuj KL}-1\right] \Big\}\right]\frac{\fz}{2k}+\cO(\fz^2).\nn
    \eea
In particular, whenever $k=K\nuj/2=2\pi\nuj m/L$ for some positive integers $\nuj$ and $m$, we have
    \begin{align}
    &R^l=\frac{-ic_{-\nuj} L^2\fz}{4\pi m\nuj}+\cO(\fz^2), &&
    R^r=\frac{-ic_\nuj L^2\fz}{4\pi m\nuj}+\cO(\fz^2), &&
    T=1-\frac{c_0L^2\fz}{4\pi m\nuj}+\cO(\fz^2),
    \label{uni-dir-3}
    \end{align}
The following theorem is a direct consequence of these equations.
    \begin{itemize}
    \item[]\textbf{Theorem~2:} Let $v:\R\to\C$ be a piecewise continuous locally periodic potential with period $\ell$ that vanishes outside $[0,L]$, and $a_n:=\int_0^\ell e^{-2\pi in x/\ell}v(x)dx$ for each integer $n$. Suppose that there are positive integers $m$ and $\nuj$ such that $L=2m\ell$ and $a_{-\nuj}=0\neq a_\nuj$ (reps.\ $a_{\nuj}=0\neq a_{-\nuj}$). Then $v$ is perturbatively unidirectionally reflectionless from the left (resp.\ right) for the wavelength $\lambda=2\ell/\nuj=L/m\nuj$. If in addition $a_0=0$, it is perturbatively unidirectionally invisible from the left (resp.\ right) for this wavelength.
	\end{itemize}
For real potentials of the form (\ref{LP-v}), we have $a_{-\nuj}=a_\nuj$, and the hypothesis of Theorem~2 does not hold. This is consistent with the fact that real potentials cannot support unidirectional invisibility \cite{pra-2013a}. In contrast, according to Theorem~2, complex potentials of the form (\ref{LP-v}) with $f(x)=\sum_{\nuj=1}^\infty c_\nuj e^{i\nuj Kx}\neq 0$ and $K$ being an integer multiple of $2\pi/L$ display perturbative unidirectional invisibility \footnote{The spectral singularities of the infinite-range analogs of these potentials are studied in \cite{gasymov}.}.

\begin{figure}
	\begin{center}
	\includegraphics[scale=.8]{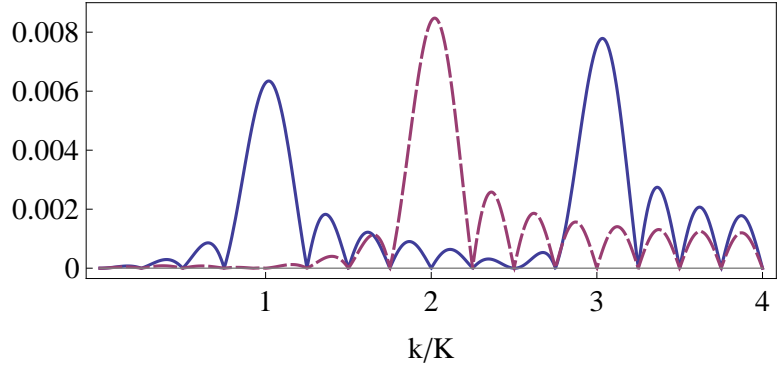}
	\caption{(Color online) Graphs of $|R^l|$ (thick navy curve), $|R^r|$ (dashed purple curve), $|T-1|$ (thin gray line) as a function of $k/K$ for a potential of the form (\ref{LP-v}) with $\fz=10^{-3}k^2$ and $f$ given by (\ref{f=34}), $\fa=2/3$, $\fb=2/5$, and $KL=4\pi$. This potential displays multimode unidirectional invisibility with $k$-dependent direction of invisibility. It is unidirectionally invisible from the left for $k=2K$ and from the right for $k=K$ and $3K$.}
	\label{fig1}
	\end{center}
	\end{figure}

Another remarkable consequence of Theorem~2 is the existence of potentials possessing perturbative unidirectional invisibility both from the left and the right for different wavenumbers. A concrete example is a potential of the form (\ref{LP-v}) with
    \be
	f(x)=e^{-2iK x}+\fa\,e^{4iK x}+\fb\,e^{-6iK x},
    \label{f=34}
	\ee
where $\fa$ and $\fb$ are nonzero complex numbers and $K$ is a real number such that $KL$ is an integer multiple of $4\pi$. This potential displays perturbative unidirectional invisibility from the left for  $k=2K$ and from the right for $k=K$ and $3K$. A graphical demonstration of this property is given in Figure~\ref{fig1} where we plot the graphs of $|R^l|$, $|R^r|$, and $|T-1|$ as a function of $k/K$ for $\fz=10^{-3}k^2$, $\fa=2/3$, $\fb=2/5$, and $KL=4\pi$. We have taken $\fz$ to be a constant multiple of $k^2$ so that the potential describes an effectively one-dimensional optically active medium with complex-refractive index $\fn:=\sqrt{1-v/k^2}$. For the above values of the parameters, $|\fn-1|<0.0012$.

Theorem~2 also foresees the extreme situations where unidirectional invisibility from the left and right occurs for infinitely many wavenumbers. A concrete example is a finite-range potential of the form (\ref{LP-v}) with $f$ given by
    \[f(x)=\sum_{j=1}^\infty \left[\fa^j e^{2ijKx}+\fb^j e^{-(2j-1)iKx}\right]
    =\frac{\fa\,e^{2iKx}}{1-\fa\,e^{2iKx}}+\frac{\fb\,e^{-iKx}}{1-\fb\,e^{-2iKx}},\]
$\fa$ and $\fb$ being nonzero complex numbers satisfying $|\fa|<1$ and $|\fb|<1$, and $K=2\pi/L$. This potential is perturbatively invisible from the left (resp.\ right) for the values of $k$ that are integer (resp.\ half-integer) multiples of $K$, i.e., $k=n K$ (resp.\ $k=(n+\frac{1}{2})K$) with $n=1,2,3,\cdots$.

\section{Perturbative Inverse Scattering}

An important consequence of Eq.~(\ref{M-pert-1}) is that we can recover the form of a $k$-independent scattering potential from the knowledge of the first-order perturbative expression for $M_{12}$ or $M_{21}$. This is because according to (\ref{M-pert-1}),
	\be
	v(x)=2\frac{d}{dx}\breve M^{(1)}_{12}(2x)=2\frac{d}{dx}
	\breve M^{(1)}_{21}(-2x),
	\label{v=M}
	\ee
where $\breve M^{(1)}_{ij}$ is the inverse Fourier transform of $M^{(1)}_{ij}$, i.e.,
	\be
	\breve M^{(1)}_{ij}(x):=\frac{1}{2\pi}\int_{-\infty}^\infty dk~ e^{ikx}M_{ij}^{(1)}(k).
	\label{inv-Fourier}
	\ee
In view of the fact that $M_{12}=R^r/T$ and $M_{21}=-R^l/T$, Eqs.~(\ref{v=M}) give a simple inverse scattering prescription. To describe this prescription we express the potential in the form
    \be
    v(x)=\fz\,f(x),
    \label{v-form}
    \ee
where $\fz$ is a coupling constant and $f$ is a possibly complex-valued function of $x$. Now, suppose that we  know the scattering data, i.e., $T$, $R^r$ (or $R^l$), as functions of $k$ and $\fz$. Then we can determine $M^{(1)}_{12}$ (reps.\ $M^{(1)}_{21}$) by keeping the lowest order term in the expansion of $R^r/T$ (reps.\ $-R^l/T$) in powers of $\fz$ and use (\ref{v=M}) and (\ref{inv-Fourier}) to recover the form of $v(x)$.

To examine how this prescription works in practice, we apply it to a rectangular barrier potential
	\be
	v(x)=\left\{
    \begin{array}{cc}
    \fz & {\rm for}~x\in(0,L),\\
    0 & {\rm otherwise}.\end{array}\right.
	\label{barrier}
	\ee
The transfer matrix for this potential is well-known \cite{pra-2013a}:
	\begin{align}
	 M_{11}(k) =M_{22}(-k)&=\left[\cos(\!\fn k L)+
	 \frac{i(\fn^2+1)\sin(\!\fn k L)}{2\fn}\right]e^{-ik L},
	\label{M11=}\\
	M_{12}(k) =M_{21}(-k)&=\frac{i(\fn^2-1)\sin(\!\fn k L) e^{-ik L}}{2\fn},
	\label{M12=}
	\end{align}
where $\fn=\sqrt{1-\fz/k^2}$. Expanding the right-hand side of (\ref{M12=}) in powers of $\fz$ and keeping the lowest order term, we find
	\be
	M_{12}^{(1)}(k) =M_{21}^{(1)}(-k)=\frac{\fz}{4k^2}\left(e^{-2i k L}-1\right).
	\label{M12=pert}
	\ee
In view of this equation and (\ref{inv-Fourier}), $\breve{M}_{12}^{(1)}(x)=\fz\left(|x|-|2L-x|\right)/8$. Substituting this relation in the first equation in (\ref{v=M}), we recover (\ref{barrier}).

As a second example, we use the above perturbative inverse scattering prescription to obtain the potential with $M^{(1)}_{12}(k)= \fz(e^{-2 i L k}-1) (e^{-i (L+J) k}-1 ) /4 k^2$, where $L,J\in\R^+$. This gives
	\be
    v(x)=\left\{\begin{array}{cc}
	-\fz & {\rm for}~0<x <L,\\[3pt]
	\fz & {\rm for}~L+J<x <2L+J,\\[3pt]
	0 &{\rm otherwise.} \end{array}\right.
    \label{v17}
    \ee
The following are two other interesting examples where we can obtain the form of the potential from that of $M^{(1)}_{12}$ analytically.
    \begin{align}
    &M^{(1)}_{12}(k)=\fz\,e^{-(Lk)^2}, && v(x)=-\frac{2\,\fz\,x}{\sqrt{\pi}L^3}\, e^{-(x/L)^2},
    \label{M-V-1}\\
    &M^{(1)}_{12}(k)=\frac{\fz}{Lk}\, e^{-(Lk)^2}, && v(x)=\frac{i\,\fz\,}{\sqrt\pi\,L^2}\,e^{-(x/L)^2}.
    \label{M-V-2}
    \end{align}
Notice that if $\fz$ is purely imaginary, the first of these potentials is $\cP\cT$-symmetric \cite{p123,PT} while the second is a real Gaussian potential.

We can formulate the above perturbative inverse scattering prescription in terms of the reflection amplitudes rather than the off-diagonal entries of the transfer matrix. In view of (\ref{RRT-pert-1}), we have
	\bea
	&&\tilde v(k)=[ik-\alpha]R^r_1(k/2),
	\label{tv=R}\\
	&&\alpha:=\tilde v(0)=\int_{-\infty}^\infty v(x)dx.
	\label{alpha=}
	\eea
Taking the inverse Fourier transform of (\ref{tv=R}) yields
	\be
	v(x)=2\left[\frac{d}{dx}-\alpha \right]\breve R^r_1(2x).
	\label{v-R}
	\ee
Now, we substitute this expression in (\ref{alpha=}). This leads to a linear equation for $\alpha$ whose solution is
	\be
	\alpha=\frac{2 \left[\breve R^r_1(\infty)-\breve R^r_1(-\infty)\right]}{1+
	\int_{-\infty}^\infty \breve R^r_1(x)dx}.
	\ee
We can determine the potential in terms of $R^r_1$ by inserting this equation in (\ref{v-R}). Pursuing the same approach we can express the potential in terms of $R^l_1$ as follows.
	\begin{align}
	&v(x)= 2\left[\frac{d}{dx}+\alpha \right]\breve R^l_1(-2x),
	&&\alpha= \frac{-2 \left[\breve R^l_1(\infty)-\breve R^l_1(-\infty)\right]}{1-
	\int_{-\infty}^\infty \breve R^l_1(x)dx}.
	\label{v=RL6}
	\end{align}
	
For situations where the first Born approximation is considered to be reliable, we can substitute $R^{r/l}$ for $R^{r/l}_1$ in (\ref{v-R}) -- (\ref{v=RL6}). This gives rise to a well-known approximate inverse scattering scheme \cite{book2}. The fact that it is the first Born approximation of the scattering data that determines the potential is elucidated in \cite{snieder}. Our two-component formulation of the scattering problem provides a simple demonstration of this surprising fact. It also highlights the possibility of its application for complex potentials.

As an illustrative example, consider the case where
	\be
	R^l_1(k)=\fz\left(\frac{k}{K}-1\right)^2e^{-L^2(k-K)^2},~~~~K,L\in\R^+.
	\label{eg01}
	\ee
Taking the inverse Fourier transform of this expression and using (\ref{v=RL6}), we find $\alpha=0$ and
	\be
	v(x)=\frac{\fz}{\sqrt\pi K^2L^7}\:
	e^{-2iKx}e^{-x^2/L^2}\left[2x^3-3L^2x+iKL^2(2x^2-L^2)\right].
    \label{inf-rang}
	\ee
According to (\ref{eg01}), this potential is perturbatively reflectionless from the left for the wavenumber $K$. To determine if it is unidirectionally invisible, we insert (\ref{inf-rang}) in (\ref{RRT-pert-1}) to compute $T_1(k)$. This gives $T_1(k)=0$. Therefore (\ref{inf-rang}) provides an example of an infinite-range scattering potential displaying perturbative unidirectional invisibility. Note that for purely imaginary values of $\fz$, it is $\cP\cT$-symmetric \cite{p123,PT}.

\section{Summary and Concluding Remarks}

The use of a two-component formulation of time-independent Schr\"odinger equation allows for expressing this equation as a time-dependent Schr\"odinger equation for a generically non-Hermitian $2\times 2$ matrix Hamiltonian $\bH(\tau)$ where $\tau$ is the product of the wavenumber $k$ and the position variable $x$. This is a rather expected and well-known fact \cite{U-M}. What is quite unexpected and remarkable is that if we choose the two-component state vectors of Refs.~\cite{jmp-1998,cqg-2003-ap-2004} for this purpose, the $S$-matrix of $\bH(\tau)$ gives the transfer matrix of $v$. A natural outcome of this observation is a straightforward derivation of a perturbative expansion for the transfer matrix $\bM$. We have used this expansion to conduct a thorough investigation of the finite-range locally periodic complex potentials displaying unidirectional reflectionlessness and invisibility. In particular, we have found a large class of generalizations of the exponential potential (\ref{exp-v}) which has served as the primary example of a potential possessing unidirectional invisibility
\cite{invisible1,invisible,U-M,invisible-old}.

Our results reveal the possibility of having multimode unidirectional invisibility with the wavelength-dependent direction of invisibility. This can indeed be achieved at finitely or infinitely many values of the wavelength. We have constructed concrete finite-range potentials with these properties.

We expect multimode unidirectional invisibility to find interesting applications in optics. For example it is tempting to envisage spectral analyzers operating on the basis of an optical potential with tunable infinitely many unidirectionally invisible modes.

Another notable consequence of studying the perturbation expansion of $\bM$ is a simple proof of the fact that the form of the potential is determined by the expression obtained in the first Born approximation for any of the off-diagonal entries of $\bM$ or equivalently the left or right reflection amplitudes. This gives rise to a perturbative inverse scattering prescription which we can also use to construct complex potentials with desired scattering properties. An interesting example is an infinite-range potential supporting perturbative unidirectional invisibility.\\

\noindent {\em Acknowledgments:} I would like to thank Aref Mostafazadeh for helpful discussions and Hugh Jones for bringing Refs.~\cite{invisible-old} to my attention. This work has been supported by  the Scientific and Technological Research Council of Turkey (T\"UB\.{I}TAK) in the framework of the project no: 112T951, and by the Turkish Academy of Sciences (T\"UBA).

\ed